# How Much do People Care about Climate Natural Disasters?

27 February 2026

Aatishya Mohanty [a], Nattavudh Powdthavee [b,d], Cheng Keat Tang [b], Andrew J. Oswald[c,d*]

[a] *University of Aberdeen, UK, AB24 3QY.*
[b] *Nanyang Technological University, Singapore, 639798.*
[c] *University of Warwick, UK, CV4 7AL.*
[d] *Wellbeing Research Centre, Harris Manchester College, Oxford OX1 3TD.*

[*]To whom any correspondence can be addressed: Telephone: +44 1437729452. Email: andrew.oswald@warwick.ac.uk. Full address: University of Warwick, Coventry CV4 7AL, United Kingdom.

ACKNOWLEDGEMENTS: For suggestions, we wish to thank Amanda Goodall, Brian Hoskins, David Stainforth, Talia Tamarin-Brodsky, and Menghan Yuan.
WORD COUNT: 4000 words approx. (excluding tables and the Appendix with supplemental material).
AUTHOR CONTRIBUTIONS: All authors designed the analysis and agreed the conclusions.
DATA AVAILABILITY: The data sets are publicly available.
AI: AI was not used in any of this work or the manuscript.
FUNDING STATEMENT: No explicit funding source.
CONFLICT OF INTEREST: The authors declare none.
ETHICS APPROVAL: Not applicable, although approval was gained by the original data-set collectors.

**Abstract**

Scientists agree about the urgency of the problem of climate change. Most citizens, however, pay little attention to gradually increasing temperature levels. Growing numbers of natural disasters in the world might then play a fundamental role as the key signal to alert humanity to the severity of the problem of the changing climate. But is that potential mechanism working? In this empirical examination (N>2 million over three decades in 93 countries), we show for the first time that a typical person's happiness and life satisfaction is barely affected by natural disasters in their region. Yet these are the individuals -- as opposed to the minority literally flooded or literally badly affected by hurricanes -- who effectively shape how governments act. This study's 'psychological near-irrelevance' result is deeply troubling.

# INTRODUCTION

More than 99.0% of scientists believe in anthropogenic climate change [1]. Yet the Earth's temperature continues to rise [2, 3]. At the time of writing, the citizens of the world are not behaving in a way that will prevent further climate change, and only 20% of Americans, for example, say that global warming is extremely personally important to them [4, Figure 12].

What explains the apparent disconnect between facts and feelings? This paper's objective is to try to explain, and offer evidence for, a potential element in an explanation. The paper's argument has three components.

First, although recent evidence suggests that higher temperatures lead to damaging outcomes for human beings, including higher numbers of suicides and greater levels of general violence [5-7], the comparatively slow and uneven increase in global temperature is itself apparently making only a limited impression on Earth's citizens. Second, given that perceptions are crucial, what is likely to be more noticeable by individuals, and thus something that might promote them to take corrective action, is the occurrence of climate-related disasters like storms, floods, and droughts [8-10]. *Disasters could, in principle, be the key signal to human beings that the planet's climate is starting to become out of control.* Third, an appropriate scientific objective is therefore to explore the extent to which average citizens' levels of psychological wellbeing are harmed by having natural disasters in their region. In many nations it is the voting behaviour of the mass of typical citizens, rather than of the unfortunate minority who literally have their homes flooded or destroyed by wildfires, who dictate policy. It is thus necessary to assess the repercussions of natural disasters beyond those people intimately affected.

We address this issue. The paper's statistical estimates -- given below -- do uncover some evidence of responsiveness of average psychological wellbeing to natural disasters. We show in large samples that wellbeing effects from disasters are negative in a statistically significant sense at the 95% confidence level. Nevertheless, in a practical sense the effect size is close to zero (the estimated elasticity lies in the interval [-0.010, 0]).

To our knowledge, no previous research study has followed the kind of multi-country methodology in the current paper. However, this paper's results are consistent in spirit with important related work on U.S. counties by Ahmadiani and Ferreira [11] who do not support "an indirect psychological impact of disaster on neighboring communities" even when those adjacent areas are in the same TV-coverage region.

More broadly, the scholarly background is familiar. There is a literature on the potential links between, especially, flooding and mental wellbeing among those directly affected [12-18]. It has been established in a large set of writings that there are empirical links between weather/climate and wellbeing [19-36].

Nevertheless, greenhouse-gas emissions in the world are rising. "With every additional increment of global warming, changes in extremes continue to become larger… Compound heatwaves and droughts are projected to become more frequent … intensification of tropical cyclones and/or extratropical storms… and increases in aridity and fire weather" IPCC Synthesis Report [2, 37]. Economists, as well as climatologists, continue to alert us [38]. It is known that attendant natural disasters, which many scientists believe will become more common with climate change, are costly in human and animal lives, health, and economic consequences. We follow partly in the innovative footsteps of researchers such as Maddison and Rehdanz [30]. Starting with Kimball et al. [27], who use Gallup data over more than a decade, it has been established that hurricane risk leads to reduced levels of life satisfaction across the areas (at zip-code level) of the USA.

**RESULTS**

To begin to test for a disasters effect, the regression equations in Tables 1-2 use international data on the World Values Survey (WVS) sample of countries and periods.

Consistent information is available for 93 nations over the period 1990-2018. Further details are described in Materials and Methods, and in [39].

Table 1 uses life satisfaction scores as a dependent variable (the scores, as answered by respondents in the survey, are coded from 0 to 10). It is possible to object to such a measure, but a large literature in social science has drawn on this form of subjective wellbeing data, and in longitudinal datasets it has been shown that cardinal satisfaction scores are a strong and almost linear predictor of the probability of individuals' subsequent actions [40]. As a complement, Table 2 uses happiness scores as its dependent variable (those, as answered by respondents in the survey, are coded on a scale from 0 to 3). Happiness can be viewed as a measure of affective well-being [41] and societal wellbeing scores have been recommended as a replacement for GDP targets [42].

Tables 1 and 2 are annual fixed-effects equations. This is to allow a judgment about the impact on people within the year of a particular natural disaster. Other independent variables -- to correct for potential confounders -- are the ones that are conventional (for example, [43-44]) in the statistical literature on the study of human-wellbeing equations: they are age, income, education, sex, marital status, health status, religiosity, employment status, and self-reported honesty.

Both Table 1 and Table 2 include a variable for the natural logarithm of the number of disasters. These are entered in the regression equations as (i) the combined total number of disasters in the fourth column of the tables and (ii) separately in columns 1, 2, and 3 as numbers of an individual type of disaster. Across the 93 countries, there are 1560 sub-regions. Year effects and sub-region effects are included on the right hand side of the equations; hence this is effectively a 28-years times 1560-regions fixed-effects approach. The Supplementary file gives Inverse Hyperbolic Sine, rather than log-form, results.

The equations reveal that natural disasters in the (broad) geographical area are associated with reduced people's feelings of life satisfaction and happiness. Each of the eight estimated coefficients in Table 1 and Table 2 is negative. In Table 1, the null of zero can be rejected at conventional confidence levels in three of four columns (floods, in column 2, being the exception). The same is true in Table 2 for happiness data. Because these are effectively fixed-effect wellbeing equations at the level of the geographical area, and given that natural disasters in a particular year can be taken as exogenously determined by the climate system, these coefficients in Tables 1 and 2 might arguably be viewed as causal.

Thus disasters have effects. However, a further, and for this paper a fundamental, issue concerns the size of the estimated consequences upon wellbeing. The disasters variables in Tables 1-2 are entered in logarithmic form. Hence it is possible to calculate what might be termed the elasticity of individuals' wellbeing with respect to disasters in their geographic area. To do that, it is necessary to divide the key coefficient by the mean of the dependent variable (approximately 6.7 in the case of life satisfaction and approximately 2.1 in the case of happiness). Consider that calculation for the final column of each of the two tables. What emerges is a small elasticity of below[1] -0.01.

Hence, for both kinds of psychological dependent variables, a doubling of the number of disasters would be associated with a drop of less than one percent in measured psychological wellbeing on a cardinal scale. That result, it should be emphasized, is for international data covering the 93 nations for which data were fully available. The coefficient's size would imply, given the mean of life satisfaction, a fall of less than 0.07 points in life satisfaction after a doubling of disasters. To put this number into perspective, the 'I am currently separated' (in a marital sense) coefficient in the life satisfaction equation is approximately -0.21.

---

[1] Technically, the numbers are actually slightly lower than implied by a standard calculation, which is why the elasticity is below -0.010. That is because, to adjust for occasional zero values, we added unity inside the Log form. The Supplementary Information file shows instead alternative Inverse Hyperbolic Sine IHS versions with similar findings.

The individual elasticity-of-life-satisfaction coefficients in columns 1-3 of Table 1 is, in each case, small. For storms, floods, and droughts, they are, respectively, -0.02, zero, and -0.04. Table 2 allows calculations for the equivalent happiness effects. The elasticities across the columns of Table 2 are -0.01, zero, -0.08, and -0.01.

Even the statistical significance of natural disasters does not hold in psychological wellbeing equations for the United States. Tables 3a-3b switch to that nation and attempt to check the WVS results using BRFSS data. The data consist of approximately 1.7 million observations from randomly sampled residents of the United States. Here the focus is once again on fixed-effects equations. In this case the regional units are U.S. states. Because of the need for data availability on wellbeing measures at the fine geographical-area level, the time period covered in the analysis is now restricted to years 2005-2011. Overall, for the United States, in the life satisfaction and happiness equations in Tables 3a-3b, it is not possible on disaster variables to reject the null of zero at standard confidence levels. The second of the tables checks the GDIS-data results of Table 3a, and finds a consistent outcome using the government Federal Emergency Management Agency (FEMA) data source on disasters.

These global and U.S. estimates fail to uncover large wellbeing effects -- at the level of geographical areas measured over a full year -- from natural disasters.

## DISCUSSION

Natural disasters are increasing. That might, in principle, be an effective signal, and perhaps the single most effective signal, to the human race that it is now essential to face the severity of the problem of climate change.

Is that signal working in practice? The broad conclusion in this paper is that the answer is approximately no. Few analysts would doubt that the individuals whose homes are literally flooded, or who experience local hurricanes, are badly affected psychologically[2]. When the level of a whole region is examined, however, this paper's calculations suggest that for typical people in a region the size of these natural-disaster psychological losses is currently tiny.

The exact finding from international data, expressed in proportional terms, and using a numerical scale for variables such as life satisfaction or happiness, is that the elasticity of humans' psychological wellbeing with respect to natural disasters is in absolute terms less than -0.010. For the U.S., we cannot reject the null hypothesis that the number is zero. Evidence on muted political reactions per se to extreme events have been examined before [45-47] and, somewhat indirectly, could be viewed as consistent with these wellbeing results.

This study's analysis implies that in international data a doubling of the number of natural disasters -- a dramatic alteration in their frequency -- in a region would be associated with a decline of less than one percentage point in a cardinalized life-satisfaction measure of human wellbeing. No matter how painful disasters are for those closely affected, it is therefore conceivable that, until natural disasters in the world are far more common than they are today, there may be relatively little support among the majority of voters for expensive strategies to contain climate change. This is a troubling implication.

---

[2] It is necessary to recall that those citizens intimately affected by the disasters are not the predominant group being measured in the analysis.

# MATERIALS AND METHODS

Information on natural disasters comes principally from the Geocoded Disasters (GDIS) dataset held at the Centre for Research on the Epidemiology of Disasters (CRED) Emergency Events Database (EM-DAT). This source reports the location of 9,924 disasters that occurred worldwide from 1960 - 2018. Natural disasters covered by the database include floods, storms (typhoons, monsoons etc.), earthquakes, landslides, droughts, volcanic activity, and extreme temperatures. Based on the latitude and longitude for each recorded disaster, we match the events to the different sub-national regions defined within the World Values Survey (WVS) data set.

A second data set on natural disasters is available, and is used later, for the United States. That information comes from the Federal Emergency Management Agency (FEMA).

The World Values Survey is a cross-country longitudinal survey that measures public attitudes, values, and beliefs on social, economic, political, and cultural subjects. This survey comprises more than 200,000 observations on randomly sampled citizens' life satisfaction and happiness, and also on people's personal characteristics and a range of socio-economic outcomes. These are collected at the individual level from 105 countries and spanning 7 waves: 1981-1984, 1989-1993, 1994-1999, 1999-2004, 2005-2009, 2010-2014, and 2017-2020.

For the later analysis, it is necessary for geo-coded climate data to be matched at the sub-national level to the survey data on individuals sampled in the WVS. We were able to achieve that for sub-national regions within a sample of 93 countries over the period 1990 to 2019. The full list of countries used in the analysis is provided below in the supplementary materials. We do the equivalent, over a shorter time span, using data from the Behavioral Risk Factor Surveillance System -- explained below -- on the counties and states of the United States.

An objective is to assess the possible psychological harm from climate-change phenomena. A WVS question, which has been used by previous researchers, asks respondents to rate their satisfaction with their lives on a scale of 1 to 10, from "dissatisfied" (lowest value on the scale) to "satisfied" (highest value). Likewise, an additional question asks respondents to indicate how happy they feel over the last one year. The responses here range from "not happy at all" (taking a value of 0) to "very happy" (taking a value of 3).

Life satisfaction data on randomly sampled American citizens is also available. This is from the Behavioral Risk Factor Surveillance System (BRFSS) database. The BRFSS is a health and behavioral risk survey conducted in the United States by the Centers for Disease Control and Prevention (CDC). It has run annually since 1984. Sample size is approximately 350,000 people per year. The survey gathers data on risk behaviors and on preventive health practices that may have an impact on an adult's health status.

In this paper we use for the United States data work the BRFSS question "In general, how satisfied are you with your life?" to construct a measure of life satisfaction. The answers are on a scale of one to four and can range from "very dissatisfied" (lowest value on the scale) to "very satisfied" (highest value). Our sample here contains slightly less than two million randomly sampled American citizens from 2439 U.S. counties during the period 2005 to 2011.

We are not able to do the American section of the analysis for a happiness question.

*Empirical Strategy*

We estimate the effect of natural disasters type $k$ ($k$ = {storm, floods and droughts}) on wellbeing outcome $h$ ($h$ = {happiness, life satisfaction}) using the following form:

$$W^h_{irt} = \alpha_r + \theta \ln(D^k_{rt}) + X'_{it}\gamma + \tau_t + \varepsilon_{irt} \qquad (1)$$

The dependent variable, $W^h_{irt}$, is self-reported happiness (Scale: 1-3) or life satisfaction (Scale: 1-10) associated with respondent $i$ residing in sub-region $r$ in year $t$. Our independent variable of interest, $ln(D^k{}_{rt})$, is the natural logarithm of the counts of disaster $k$ in region $r$ at year $t$ and thus $\theta$ can be thought of as a measure of the change in well-being outcomes from a 1% increase in disaster counts. We estimate a total of 6 different regressions for the different disaster types on the different well-being outcomes.

We also add unity inside the log form. The mean number of regions with zero disasters every year is approximately 60 (in the cross-country dataset).

Because of the possibility of spatial autocorrelation, Conley standard errors are presented in Tables 1 and 2. Tables 3a,b have insignificant coefficients even without Conley standard errors.

In the analysis we are effectively calculating the consequences of natural disasters for citizens' well-being with a conventional fixed-effects approach. Specifically, we augment sub-national geographical area fixed effects ($\alpha_r$) and year fixed effects ($\tau_t$) to minimize the risk of time-invariant unobserved differences across sub-national areas, and general trends in well-being outcomes from biasing our estimates. We further include a set of observable respondents' characteristics (age, income, education, sex, marital status, health status etc.) that are standard in the empirical literature on wellbeing in order to minimize the risk of problems from omitted confounding variables.

The above specifications imply that the estimated structures are reduced-form ones (as has been true for almost all previous social-science research in the area of climate-change research). That has some disadvantages. A variety of questions, about the nature of possible transmission mechanisms, remain unanswered in reduced-form econometric work. Nevertheless, such an approach is flexible and appears to be appropriate at the current juncture in research.

**Table 1. Cross-National Life Satisfaction Equations with Disaster Variables – World Values Survey Data 1990 to 2018**

|  | (1) | (2) | (3) | (4) |
|---|---|---|---|---|
| Storm (Log) | -0.131** [-0.232,-0.030] | | | |
| Flood (Log) | | -0.032 [-0.116,0.052] | | |
| Drought (Log) | | | -0.247** [-0.491,-0.003] | |
| Total disasters (Log) (i.e. their sum) | | | | -0.078** [-0.149,-0.006] |
| Observations | 257230 | 257230 | 257230 | 257230 |
| R-squared | 0.12 | 0.12 | 0.12 | 0.12 |
| No. of countries | 93 | 93 | 93 | 93 |

Sub-national region fixed effects and year dummies are included in all columns. The number of subnational regions is 1560. 'Log' here, and in later tables, is the natural logarithm.

The dependent variable is the WVS question "Rate your life satisfaction" (Scale: 1-10). Storms, floods, droughts, and total disasters measure the annual number of occurrences in GDIS data of each disaster in each sub-national region. GDIS (Geocoded Disasters) data is an open-source extension to the EM-DAT database, providing spatial information (geolocations) for natural disasters, formally, from the Centre for Research on the Epidemiology of Disasters (CRED)'s EM-DAT (Emergency Events Database).

Conley standard errors at the sub-national region level, at 500km, are used to calculate 95% confidence intervals as reported in square brackets. ***, **, and * denote significance level at 1%, 5%, & 10%, respectively. The baseline controls include age dummies, income, education, sex, marital status, health status, religiosity, employment status and self-reported honesty.

The mean of the dependent variable is 6.67. Its standard deviation is 2.39.

**Table 2. Cross-National Happiness Equations with Disaster Variables – World Values Survey Data 1990 to 2018**

|  | (1) | (2) | (3) | (4) |
|---|---|---|---|---|
| Storm (Log) | -0.030* [-0.065,0.005] | | | |
| Flood (Log) | | -0.011 [-0.034,0.011] | | |
| Drought (Log) | | | -0.167*** [-0.280,-0.055] | |
| Total disasters (Log) (i.e. their sum) | | | | -0.029*** [-0.050,-0.007] |
| Observations | 258113 | 258113 | 258113 | 258113 |
| R-squared | 0.15 | 0.15 | 0.15 | 0.15 |
| No. of countries | 93 | 93 | 93 | 93 |

Sub-national region fixed effects and year dummies are included in all columns. The number of subnational regions is 1560.

The dependent variable is the WVS question "How happy are you" (Scale: 0-3). Storms, floods, droughts, and total disasters measure the annual number of occurrences in GDIS data of each disaster in each sub-national region.

Conley standard errors at the sub-national region level, at 500km, are used to calculate 95% confidence intervals as reported in square brackets. ***, **, and * denote significance level at 1%, 5%, & 10%, respectively. The baseline controls include age dummies, income, education, sex, marital status, health status, religiosity, employment status and self-reported honesty.

The mean of the dependent variable is 2.09. Its standard deviation is 0.74.

**Table 3a. U.S. Life Satisfaction Equations with Disaster Variables – United States Data 2005 to 2011 [GDIS data]**

|                     | (1)             | (2)              | (3)             | (4)              |
|---------------------|-----------------|------------------|-----------------|------------------|
| Storm (Log)         | 0.002           |                  |                 |                  |
|                     | [-0.003,0.007]  |                  |                 |                  |
| Flood (Log)         |                 | -0.005           |                 |                  |
|                     |                 | [-0.015,0.005]   |                 |                  |
| Drought (Log)       |                 |                  | 0.009           |                  |
|                     |                 |                  | [-0.010,0.028]  |                  |
| Total disasters (Log)|                |                  |                 | -0.0001          |
|                     |                 |                  |                 | [-0.005,0.005]   |
| Observations        | 1745570         | 1745570          | 1745570         | 1745570          |
| R-squared           | 0.18            | 0.18             | 0.18            | 0.18             |
| No. of states       | 49              | 49               | 49              | 49               |

State fixed effects and year dummies are included in all columns.

The dependent variable is the BRFSS question "In general, how satisfied are you with your life?" (Scale: 1-4). Storms, floods, droughts, and total disasters measure the annual number of occurrences in GDIS data of each disaster in each state. The sample excludes Alaska, Hawaii, and US overseas territories. Standard errors are clustered at the state level, and 95% confidence intervals are reported in parentheses. ***, **, and * denote significance level at 1%, 5%, & 10%, respectively. The baseline controls include age dummies, income, education, sex, marital status, health status, and employment status.

The mean of the dependent variable is 3.39. Its standard deviation is 0.63.

**Table 3b - U.S. Life Satisfaction Equations with Disaster Variables – United States Data 2005 to 2011 [Federal Emergency Management Agency, FEMA, data]**

|                      | (1)             | (2)              | (3)             |
|----------------------|-----------------|------------------|-----------------|
| Storm (Log)          | -0.003          |                  |                 |
|                      | [-0.008,0.002]  |                  |                 |
| Flood (Log)          |                 | -0.009           |                 |
|                      |                 | [-0.025,0.008]   |                 |
| Total disasters (Log)|                 |                  | -0.003          |
|                      |                 |                  | [-0.008,0.002]  |
| Observations         | 1746020         | 1746020          | 1746020         |
| R-squared            | 0.18            | 0.18             | 0.18            |
| No. of states        | 49              | 49               | 49              |

State fixed effects and year dummies are included in all columns.

The dependent variable is the BRFSS question "In general, how satisfied are you with your life?" (Scale: 1-4). Storms, floods, droughts, and total disasters measure the annual number of occurrences in FEMA data of each disaster in each state. The sample excludes Alaska, Hawaii, and US overseas territories. Standard errors are clustered at the state level, and 95% confidence intervals are reported in parentheses. ***, **, and * denote significance level at 1%, 5%, & 10%, respectively. The baseline controls include age dummies, income, education, sex, marital status, health status, and employment status.

The mean of the dependent variable is 3.39. Its standard deviation is 0.63.

**Figure 3b footnote:**

Source for the USA disasters data in the FEMA database*:

Federal Emergency Management Agency (2021). Open FEMA Dataset, *Retrieved from: https://www.fema.gov/disaster*

**Figure 1:** Global Total Disasters

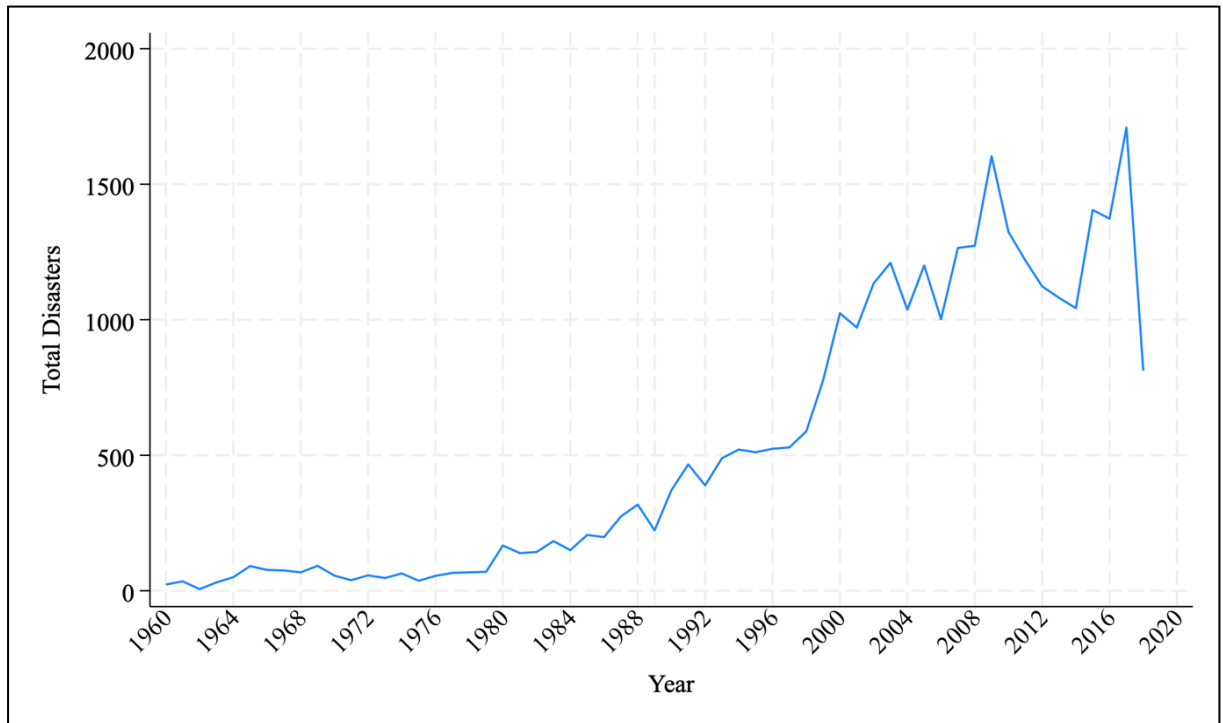

Total disasters measure the annual total number of disasters. Source: GDIS data as above.

**SUPPLEMENTARY INFORMATION FILE**
**(to be published separately)**

**Table S1A. Cross-National Life Satisfaction Equations with Disaster Variables – World Values Survey Data 1990 to 2018. Inverse Hyperbolic Sine IHS Version.**

|  | (1) | (2) | (3) | (4) |
|---|---|---|---|---|
| Storm (IHS) | -0.101** [-0.179,-0.023] | | | |
| Flood (IHS) | | -0.025 [-0.091,0.041] | | |
| Drought (IHS) | | | -0.193** [-0.384,-0.003] | |
| Total disasters (IHS) (i.e. their sum) | | | | -0.060** [-0.116,-0.004] |
| Observations | 257230 | 257230 | 257230 | 257230 |
| R-squared | 0.12 | 0.12 | 0.12 | 0.12 |
| No. of countries | 93 | 93 | 93 | 93 |

Sub-national region fixed effects and year dummies are included in all columns. The number of subnational regions is 1560. 'IHS' here, and in later tables, is the inverse hyperbolic sine.

The dependent variable is the WVS question "Rate your life satisfaction" (Scale: 1-10). Storms, floods, droughts, and total disasters measure the annual number of occurrences in GDIS data of each disaster in each sub-national region. GDIS (Geocoded Disasters) data is an open-source extension to the EM-DAT database, providing spatial information (geolocations) for natural disasters, formally, from the Centre for Research on the Epidemiology of Disasters (CRED)'s EM-DAT (Emergency Events Database).

Conley standard errors at the sub-national region level, at 500km, are used to calculate 95% confidence intervals as reported in square brackets. ***, **, and * denote significance level at 1%, 5%, & 10%, respectively. The baseline controls include age dummies, income, education, sex, marital status, health status, religiosity, employment status and self-reported honesty.

The mean of the dependent variable is 6.67. Its standard deviation is 2.39.

**Table S2A. Cross National Happiness Equations with Disaster Variables – World Values Survey Data 1990 to 2018. Inverse Hyperbolic Sine IHS Version.**

|  | (1) | (2) | (3) | (4) |
|---|---|---|---|---|
| Storm (IHS) | -0.023* | | | |
|  | [-0.050,0.004] | | | |
| Flood (IHS) | | -0.009 | | |
|  | | [-0.026,0.009] | | |
| Drought (IHS) | | | -0.131*** | |
|  | | | [-0.219,-0.043] | |
| Total disasters (IHS) (i.e. their sum) | | | | -0.022** |
|  | | | | [-0.039,-0.005] |
| Observations | 258113 | 258113 | 258113 | 258113 |
| R-squared | 0.15 | 0.15 | 0.15 | 0.15 |
| No. of countries | 93 | 93 | 93 | 93 |

Sub-national region fixed effects and year dummies are included in all columns. The number of subnational regions is 1560.

The dependent variable is the WVS question "How happy are you" (Scale: 0-3). Storms, floods, droughts, and total disasters measure the annual number of occurrences in GDIS data of each disaster in each sub-national region.

Conley standard errors at the sub-national region level, at 500km, are used to calculate 95% confidence intervals as reported in square brackets. ***, **, and * denote significance level at 1%, 5%, & 10%, respectively. The baseline controls include age dummies, income, education, sex, marital status, health status, religiosity, employment status and self-reported honesty.

The mean of the dependent variable is 2.09. Its standard deviation is 0.74.

**Figure S1:** Total Disasters in US states – GDIS dataset

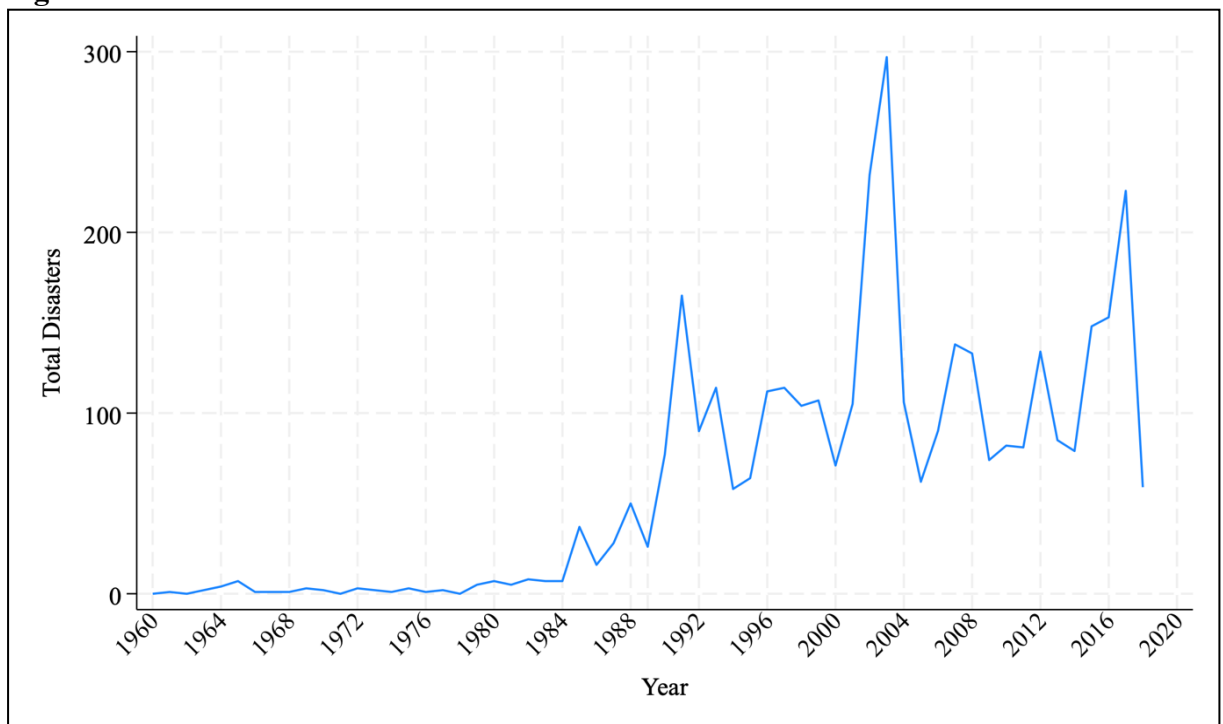

Total disasters measure the annual total number of disasters in the 49 states of our US sample using the GDIS dataset.

**Figure S2:** Total Disasters in US states – FEMA dataset

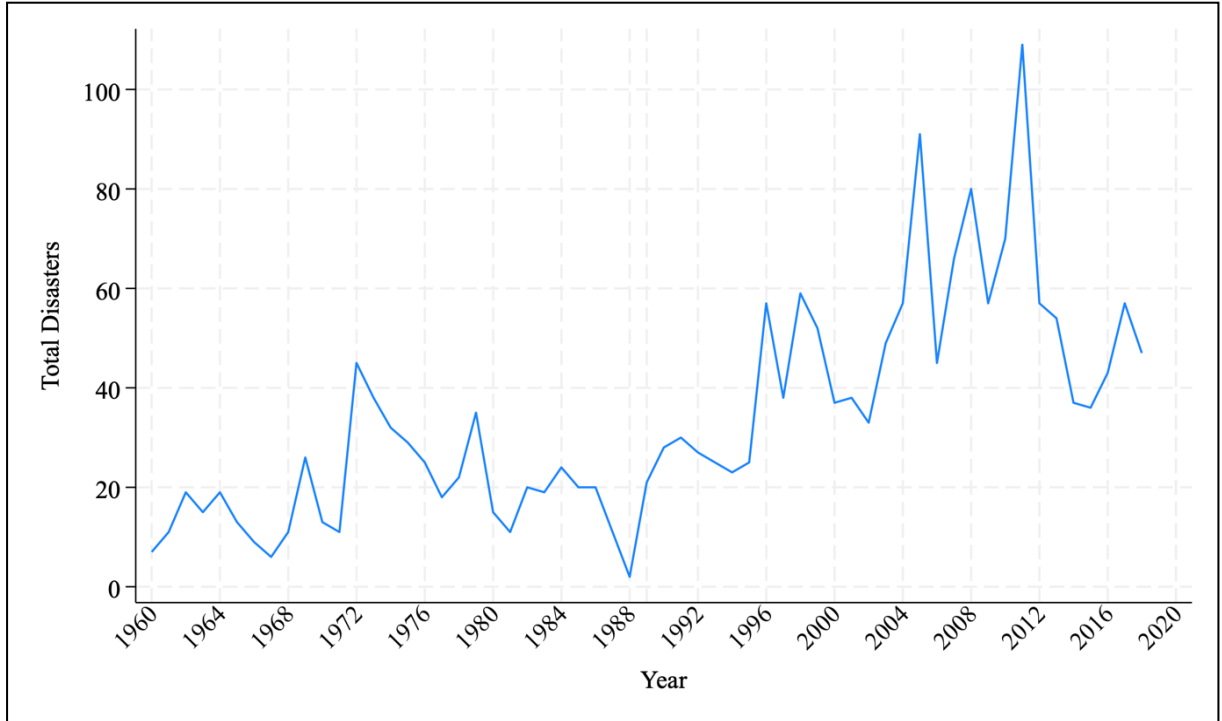

Total disasters measure the annual total number of disasters in the 49 states of our US sample using the FEMA dataset.

**Countries**

List of countries in the WVS sample

1. Albania
2. Algeria
3. Argentina
4. Armenia
5. Australia
6. Azerbaijan
7. Bangladesh
8. Belarus
9. Bolivia
10. Bosnia and Herzegovina
11. Brazil
12. Bulgaria
13. Burkina Faso
14. Canada
15. Chile
16. China
17. Colombia
18. Cyprus
19. Czech Republic
20. Dominican Republic
21. Ecuador
22. Egypt
23. El Salvador
24. Estonia
25. Ethiopia
26. Finland
27. France
28. Georgia
29. Germany
30. Ghana
31. Greece
32. Guatemala
33. Haiti
34. Hong Kong
35. Hungary
36. India
37. Indonesia
38. Iran
39. Iraq
40. Italy
41. Japan
42. Jordan
43. Kazakhstan
44. Kuwait
45. Kyrgyzstan
46. Latvia
47. Lebanon
48. Libya
49. Lithuania
50. Malaysia
51. Mali
52. Mexico
53. Moldova
54. Montenegro
55. Morocco
56. Netherlands
57. New Zealand
58. Nigeria
59. North Macedonia
60. Norway
61. Pakistan
62. Palestine
63. Peru
64. Philippines
65. Poland
66. Puerto Rico
67. Romania
68. Russia
69. Rwanda
70. Serbia
71. Slovakia
72. Slovenia
73. South Africa
74. South Korea
75. Spain
76. Sweden
77. Switzerland
78. Taiwan
79. Thailand
80. Trinidad and Tobago
81. Tunisia
82. Turkey
83. Uganda
84. Ukraine
85. United Kingdom
86. United States
87. Uruguay
88. Uzbekistan
89. Venezuela
90. Vietnam
91. Yemen
92. Zambia
93. Zimbabwe

**Data Description for WVS**

| Variable | Description | Source |
|---|---|---|
| | *Main Variables* | |
| Satisfied | A measure of well-being based on the WVS question on how satisfied the respondent is with their life. The scale ranges from zero to ten with larger values corresponding to a higher level of satisfaction. | WVS Database (2020) |
| Happy | A measure of well-being based on the WVS question on how happy is the respondent. The answers can range from (0) not "happy at all" to (3) "very happy". | WVS Database (2020) |
| | *Baseline Controls* | |
| Age | The age of the respondent. | WVS Database (2020) |
| Income | The level of income of the respondent. The scale ranges from zero to ten with larger values corresponding to a higher step in the income scale. | WVS Database (2020) |
| Education | The variable measures the highest level of education attained by the respondent. The scale has three categories: low, medium and high (1-3). | WVS Database (2020) |
| Sex | The gender of the respondent. A value of one is assigned for a male and zero for female. | WVS Database (2020) |
| Marital status | Marital status of the respondent. A dummy variable is generated for each of the categories of classification - married, cohabiting, divorced, separated, widowed and single. | WVS Database (2020) |
| Health status | The variable measures the subjective health of the respondent. A dummy variable is generated for each of the categories of classification – very poor, poor, normal, good. | WVS Database (2020) |
| Religiosity | A measure for religiosity based on the WVS question on how important is religion in the respondent's life. The variable is recoded such that the answers range from (0) "not at all important" to (4) "very important". | WVS Database (2020) |
| Employment status | A dummy variable to indicate the employment status of the respondent. A value of one is assigned if employed, and zero otherwise. | WVS Database (2020) |
| Honesty | A measure for honesty based on the WVS question if it is justifiable to cheat on taxes. The variable is recoded such that the answers range from (0) "always" to (10) "Never". | WVS Database (2020) |
| | *Natural Disasters* | |
| Drought | The total annual occurrence of droughts in each subnational region. | Rosvold and Buhaug (2021) |
| Flood | The total annual occurrence of floods in each subnational region. | Rosvold and Buhaug (2021) |
| Storm | The total annual occurrence of storms in each subnational region. | Rosvold and Buhaug (2021) |

| Variable | Description | Source |
|---|---|---|
| Total Disasters | The total annual occurrence of droughts, floods, and storms in each subnational region. | Rosvold and Buhaug (2021) |

Data Description for BRFSS

| Variable | Description | Source |
|---|---|---|
| | *Main Variables* | |
| Life Satisfaction | A measure of well-being based on the BRFSS question "In general, how satisfied are you with your life?". The scale ranges from one to four with larger values corresponding to a higher level of satisfaction. | CDC (2011) |
| | *Baseline Controls* | |
| Age | The age of the respondent. | CDC (2011) |
| Income | The level of income of the respondent. The scale ranges from zero to eight with larger values corresponding to a higher step in the income scale. | CDC (2011) |
| Education | The variable measures the highest level of education attained by the respondent. The scale has six levels with increasing level of education corresponding to a higher value. | CDC (2011) |
| Sex | The gender of the respondent. A dummy variable is generated for males. | CDC (2011) |
| Marital status | Marital status of the respondent. A dummy variable is generated for each of the categories of classification - married, divorced, separated, and widowed. | CDC (2011) |
| Health status | The variable measures the subjective health of the respondent. A dummy variable is generated for each of the categories of classification – poor, fair, very good, excellent. | CDC (2011) |
| Employment status | A dummy variable to indicate the employment status of the respondent. A value of one is assigned if employed, and zero otherwise. | CDC (2011) |
| | *Natural Disasters* | |
| Drought | The total annual occurrence of droughts in each county. | Rosvold and Buhaug (2021) |
| Flood | The total annual occurrence of floods in each county. | Rosvold and Buhaug (2021) |
| Storm | The total annual occurrence of storms in each county. | Rosvold and Buhaug (2021) |
| Total Disasters | The total annual occurrence of droughts, floods, and storms in each county. | Rosvold and Buhaug (2021) |

**References for Data**

CDC (2011). Behavioral Risk Factor Surveillance System Survey Data (2005-2011). *U.S. Department of Health and Human Services, Centers for Disease Control and Prevention.*

WVS Database, (2020). World Values Survey: All Rounds Country-Pooled Datafile (eds.). *JD Systems Institute & World Values Survey Association.*